\documentstyle[12pt]{article}
\newcommand{\nc}{\newcommand}
\nc{\renc}{\renewcommand}

\nc{\etal}{\mbox{\it et al. }}
\nc{\ie}{{\it i.e.}}
\nc{\eg}{{\it e.g.}}

\renc{\thefootnote}{\arabic{footnote}}
\nc{\capt}[1]{{\bf Figure.} {\small\sl #1}}

\nc{\eqs}[2]{\mbox{Eqs.~(\ref{#1},\,\ref{#2})}}
\nc{\eq}[1]{\mbox{Eq.~(\ref{#1})}}

\nc{\figs}[2]{\mbox{Figs.~(\ref{#1},\,\ref{#2})}}
\nc{\fig}[1]{\mbox{Fig~.(\ref{#1})}}

\nc{\tag}[1]{\label{#1} \marginpar{{\footnotesize #1}}}
\nc{\mtag}[1]{\label{#1} \mbox{\marginpar{{\footnotesize #1}}}}
\renc{\baselinestretch}{1.2}
\newlength{\overeqskip}
\newlength{\undereqskip}
\nc{\be}[1]{\begin{equation} \mbox{$\label{#1}$}}
\nc{\bea}[1]{\begin{eqnarray} \mbox{$\label{#1}$}}
\nc{\Section}[2]{\section{#2}\label{#1}}
\nc{\Bibitem}[1]{\bibitem{#1}}
\nc{\Label}[1]{\label{#1}}

\nc{\eea}{\vspace{\undereqskip}\end{eqnarray}}
\nc{\ee}{\vspace{\undereqskip}\end{equation}}
\nc{\bdm}{\begin{displaymath}}
\nc{\edm}{\end{displaymath}}
\nc{\dpsty}{\displaystyle}
\nc{\bc}{\begin{center}}
\nc{\ec}{\end{center}}
\nc{\ba}{\begin{array}}
\nc{\ea}{\end{array}}
\nc{\bab}{\begin{abstract}}
\nc{\eab}{\end{abstract}}
\nc{\btab}{\begin{tabular}}
\nc{\etab}{\end{tabular}}
\nc{\bit}{\begin{itemize}}
\nc{\eit}{\end{itemize}}
\nc{\ben}{\begin{enumerate}}
\nc{\een}{\end{enumerate}}
\nc{\bfig}{\begin{figure}}
\nc{\efig}{\end{figure}}
\nc{\arreq}{&\!=\!&}
\nc{\arrmi}{&\!-\!&}
\nc{\arrpl}{&\!+\!&}
\nc{\arrap}{&\!\!\!\approx\!\!\!&}
\nc{\non}{\nonumber\\*}
\nc{\align}{\!\!\!\!\!\!\!\!&&}

\def\lsim{\; \raise0.3ex\hbox{$<$\kern-0.75em
      \raise-1.1ex\hbox{$\sim$}}\; }
\def\gsim{\; \raise0.3ex\hbox{$>$\kern-0.75em
      \raise-1.1ex\hbox{$\sim$}}\; }
\nc{\DOT}{\hspace{-0.08in}{\bf .}\hspace{0.1in}}
\nc{\Laada}{\hbox {$\sqcap$ \kern -1em $\sqcup$}}
\nc\loota{{\scriptstyle\sqcap\kern-0.55em\hbox{$\scriptstyle\sqcup$}}}
\nc\Loota{{\sqcap\kern-0.65em\hbox{$\sqcup$}}}
\nc\laada{\Loota}
\nc{\qed}{\hskip 3em \hbox{\BOX} \vskip 2ex}

\nc{\real}{{\rm I \! R}}
\nc{\Z}{{\sf Z \!\!\! Z}}
\nc{\complex}{{\rm C\!\!\! {\sf I}\,\,}}
\def\bigid{\leavevmode\hbox{\small1\kern-3.8pt\normalsize1}}
\def\id{\leavevmode\hbox{\small1\kern-3.3pt\normalsize1}}
\nc{\slask}{\!\!\!/}
\nc{\bis}{{\prime\prime}}
\nc{\pa}{\partial}
\nc{\na}{\nabla}
\nc{\ra}{\rangle}
\nc{\la}{\langle}
\nc{\goto}{\rightarrow}
\nc{\swap}{\leftrightarrow}

\nc{\EE}[1]{ \mbox{$\cdot10^{#1}$} }
\nc{\abs}[1]{\left|#1\right|}
\nc{\at}[2]{\left.#1\right|_{#2}}
\nc{\norm}[1]{\|#1\|}
\nc{\abscut}[2]{\Abs{#1}_{\scriptscriptstyle#2}}
\nc{\vek}[1]{{\rm\bf #1}}
\nc{\integral}[2]{\int\limits_{#1}^{#2}}
\nc{\inv}[1]{\frac{1}{#1}}
\nc{\dd}[2]{{{\partial #1}\over{\partial #2}}}
\nc{\ddd}[2]{{{{\partial}^2 #1}\over{\partial {#2}^2}}}
\nc{\dddd}[3]{{{{\partial}^2 #1}\over
	{\partial #2 \partial #3}}}
\nc{\dder}[2]{{{d #1}\over{d #2}}}
\nc{\ddder}[2]{{{d^2 #1}\over{d {#2}^2}}}
\nc{\dddder}[3]{{d^2 #1}\over
	{d #2 d #3}}
\nc{\dx}[1]{d\,^{#1}x}
\nc{\dy}[1]{d\,^{#1}y}
\nc{\dz}[1]{d\,^{#1}z}
\nc{\dl}[1]{\frac{d\,^{#1}l}{(2\pi)^{#1}}}
\nc{\dk}[1]{\frac{d\,^{#1}k}{(2\pi)^{#1}}}
\nc{\dq}[1]{\frac{d\,^{#1}q}{(2\pi)^{#1}}}

\nc{\cc}{\mbox{$c.c.$ }}
\nc{\hc}{\mbox{$h.c.$ }}
\nc{\cf}{cf.\ }
\nc{\erfc}{{\rm erfc}}
\nc{\Tr}{{\rm Tr\,}}
\nc{\tr}{{\rm tr\,}}
\nc{\pol}{{\rm pol}}
\nc{\sign}{{\rm sign}}
\nc{\bfT}{{\bf T }}

\nc{\cA}{{\cal A}}
\nc{\cB}{{\cal B}}
\nc{\cD}{{\cal D}}
\nc{\cE}{{\cal E}}
\nc{\cG}{{\cal G}}
\nc{\cH}{{\cal H}}
\nc{\cL}{{\cal L}}
\nc{\cO}{{\cal O}}
\nc{\cT}{{\cal T}}
\nc{\cN}{{\cal N}}
\nc{\rvac}[1]{|{\cal O}#1\rangle}
\nc{\lvac}[1]{\langle{\cal O}#1|}
\nc{\rvacb}[1]{|{\cal O}_\beta #1\rangle}
\nc{\lvacb}[1]{\langle{\cal O}_\beta #1 |}
\nc{\bb}{\bar{\beta}}
\nc{\bt}{\tilde{\beta}}
\nc{\ctH}{\tilde{\cal H}}
\nc{\chH}{\hat{\cal H}}
\nc{\1}{\aa}
\nc{\2}{\"{a}}
\nc{\3}{\"{o}}
\nc{\4}{\AA}
\nc{\5}{\"{A}}
\nc{\6}{\"{O}}
\nc{\al}{\alpha}
\nc{\g}{\gamma}
\nc{\Del}{\Delta}
\nc{\e}{\epsilon}
\nc{\eps}{\epsilon}
\nc{\lam}{\lambda}
\nc{\om}{\omega}
\nc{\Om}{\Omega}
\nc{\ve}{\varepsilon}
\nc{\mn}{{\mu\nu}}
\nc{\k}{\kappa}
\nc{\vp}{\varphi}

\nc{\advp}[3]{{\it  Adv.\ in\ Phys.\ }{{\bf #1} {(#2)} {#3}}}
\nc{\annp}[3]{{\it  Ann.\ Phys.\ (N.Y.)\ }{{\bf #1} {(#2)} {#3}}}
\nc{\apl}[3]{{\it  Appl. Phys. Lett. }{{\bf #1} {(#2)} {#3}}}
\nc{\apj}[3]{{\it  Ap.\ J.\ }{{\bf #1} {(#2)} {#3}}}
\nc{\apjl}[3]{{\it  Ap.\ J.\ Lett.\ }{{\bf #1} {(#2)} {#3}}}
\nc{\app}[3]{{\it Astropart.\ Phys.\ }{{\bf #1} {(#2)} {#3}}}
\nc{\cmp}[3]{{\it  Comm.\ Math.\ Phys.\ }{{ \bf #1} {(#2)} {#3}}}
\nc{\cqg}[3]{{\it  Class.\ Quant.\ Grav.\ }{{\bf #1} {(#2)} {#3}}}
\nc{\epl}[3]{{\it  Europhys.\ Lett.\ }{{\bf #1} {(#2)} {#3}}}
\nc{\ijmp}[3]{{\it Int.\ J.\ Mod.\ Phys.\ }{{\bf #1} {(#2)} {#3}}}
\nc{\ijtp}[3]{{\it Int.\ J.\ Theor.\ Phys.\ }{{\bf #1} {(#2)} {#3}}}
\nc{\jmp}[3]{{\it  J.\ Math.\ Phys.\ }{{ \bf #1} {(#2)} {#3}}}
\nc{\jpa}[3]{{\it  J.\ Phys.\ A\ }{{\bf #1} {(#2)} {#3}}}
\nc{\jpc}[3]{{\it  J.\ Phys.\ C\ }{{\bf #1} {(#2)} {#3}}}
\nc{\jap}[3]{{\it J.\ Appl.\ Phys.\ }{{\bf #1} {(#2)} {#3}}}
\nc{\jpsj}[3]{{\it J.\ Phys.\ Soc.\ Japan\ }{{\bf #1} {(#2)} {#3}}}
\nc{\lmp}[3]{{\it Lett.\ Math.\ Phys.\ }{{\bf #1} {(#2)} {#3}}}
\nc{\mpl}[3]{{\it  Mod.\ Phys.\ Lett.\ }{{\bf #1} {(#2)} {#3}}}
\nc{\ncim}[3]{{\it  Nuov.\ Cim.\ }{{\bf #1} {(#2)} {#3}}}
\nc{\np}[3]{{\it  Nucl.\ Phys.\ }{{\bf #1} {(#2)} {#3}}}
\nc{\pr}[3]{{\it Phys.\ Rev.\ }{{\bf #1} {(#2)} {#3}}}
\nc{\pra}[3]{{\it  Phys.\ Rev.\ A\ }{{\bf #1} {(#2)} {#3}}}
\nc{\prb}[3]{{\it  Phys.\ Rev.\ B\ }{{{\bf #1} {(#2)} {#3}}}}
\nc{\prc}[3]{{\it  Phys.\ Rev.\ C\ }{{\bf #1} {(#2)} {#3}}}
\nc{\prd}[3]{{\it  Phys.\ Rev.\ D\ }{{\bf #1} {(#2)} {#3}}}
\nc{\prl}[3]{{\it Phys\ Rev.\ Lett.\ }{{\bf #1} {(#2)} {#3}}}
\nc{\pl}[3]{{\it  Phys.\ Lett.\ }{{\bf #1} {(#2)} {#3}}}
\nc{\prep}[3]{{\it Phys\. Rep.\ }{{\bf #1} {(#2)} {#3}}}
\nc{\prsl}[3]{{\it Proc.\ R.\ Soc.\ London\ }{{\bf #1} {(#2)} {#3}}}
\nc{\ptp}[3]{{\it  Prog.\ Theor.\ Phys.\ }{{\bf #1} {(#2)} {#3}}}
\nc{\ptps}[3]{{\it  Prog\ Theor.\ Phys.\ suppl.\ }{{\bf #1} {(#2)} {#3}}}
\nc{\physa}[3]{{\it  Physica\ A\ }{{\bf #1} {(#2)} {#3}}}
\nc{\physb}[3]{{\it  Physica\ B\ }{{\bf #1} {(#2)} {#3}}}
\nc{\phys}[3]{{\it Physica\ }{{\bf #1} {(#2)} {#3}}}
\nc{\rmp}[3]{{\it  Rev.\ Mod.\ Phys.\ }{{\bf #1} {(#2)} {#3}}}
\nc{\rpp}[3]{{\it Rep.\ Prog.\ Phys.\ }{{\bf #1} {(#2)} {#3}}}
\nc{\sjnp}[3]{{\it Sov.\ J.\ Nucl.\ Phys.\ }{{\bf #1} {(#2)} {#3}}}
\nc{\spjetp}[3]{{\it Sov.\ Phys.\ JETP\ }{{\bf #1} {(#2)} {#3}}}
\nc{\yf}[3]{{\it Yad.\ Fiz.\ }{{\bf #1} {(#2)} {#3}}}
\nc{\zetp}[3]{{\it Zh.\ Eksp.\ Teor.\ Fiz.\  }{{\bf #1}  {(#2)} {#3}}}
\nc{\zp}[3]{{\it Z.\ Phys.\ }{{\bf #1} {(#2)} {#3}}}
\nc{\ibid}[3]{{\sl ibid.\ }{{\bf #1} {#2} {#3}}}
\nc{\rf}[1]{(\ref{#1})}
\nc{\nn}{\nonumber \\*}
\nc{\bfB}{\bf{B}}
\nc{\bfv}{\bf{v}}
\nc{\bfx}{\bf{x}}
\nc{\bfy}{\bf{y}}
\nc{\vx}{\vec{x}}
\nc{\vy}{\vec{y}}
\nc{\oB}{\overline{B}}
\nc{\oI}{\overline{I}}
\nc{\oR}{\overline{R}}
\nc{\rar}{\rightarrow}
\nc{\ti}{\times}
\nc{\slsh}{\hskip-5pt/}
\nc{\sm}{Standard~Model~}
\nc{\MP}{M_{\rm Pl}}
\nc{\tp}{t_{\rm Pl}}
\nc{\ave}{\bar{E}}

\renc{\min}{p_{\rm min}}
\renc{\max}{p_{\rm max}}
\nc{\pmin}{p_{\rm min}}
\nc{\pmax}{p_{\rm max}}
\nc{\fo}{f_0}
\nc{\foi}{f_{0,i}\,}
\nc{\fop}{f_0^P}
\nc{\fou}{f_0^U}
\def\sepand{\rule{14cm}{0pt}\and}
\nc{\eff}{{\rm eff}}
\nc{\MT}{M_{\rm T}}
\nc{\ML}{M_{\rm L}}
\nc{\kk}{\vek{k}}
\nc{\pp}{{\rm p}}
\nc{\cb}{critical bubble~}
\nc{\cbs}{critical bubbles~}
\nc{\scb}{subcritical bubble~}
\nc{\scbs}{subcritical bubbles~}
\begin{document}

{\title{{\hfill {{\small HU-TFT-94-41\\
         \hfill TURKU-FL-P13\\
	 \hfill hep-ph/yymmxxx
        }}\vskip 1truecm}
{\bf Induced nucleation in weak first order phase transitions}}


\author{
{\sc Kari Enqvist$^{1}$ } \\
{\sl Research Institute for Theoretical Physics} \\
{\sl P.O. Box 9, FIN-00014 University of Helsinki, Finland}\\
and\\
{\sc Iiro Vilja$^{2}$ }\\
{\sl Department of Physics,
University of Turku} \\
{\sl FIN-20500 Turku, Finland} \\
\sepand
}
\maketitle}
\vspace{2cm}
\begin{abstract}
\noindent We study induced nucleation by considering the accumulation rate
of shrinking subcritical bubbles. We derive the probability for
a collection of subcritical
bubbles to form a critical bubble, and argue that this mechanism could
well play a role in electroweak phase transitions if the Higgs is heavy.

\end{abstract}
\vfill
\footnoterule
{\small  $^1$enqvist@phcu.helsinki.fi; $^2$vilja@utu.fi}
\thispagestyle{empty}
\newpage
\setcounter{page}{1}

Cosmological phase transitions have recently received much attention,
in particular in the context of electroweak baryogenesis \cite{bg}.
First order phase transition and bubble dynamics in the Standard
Model have been studied in detail, and it has become increasingly
clear \cite{studies1, studies2, FodorH} that for realistic
Higgs masses, much heavier than 60 GeV,
the electroweak phase transition appears to be only weakly
first order. If the Higgs mass $m_H\gsim 100$ GeV, both lattice studies
and perturbative calculations run into technical troubles. It is however
conceivable that for such Higgs masses the electroweak transition is
close to a second order one, and that it proceeds not by critical bubble
formation but via thermal fluctuations.
 These would give rise to a considerable
phase mixing at the critical temperature.

In the context of the EW
phase transition,
subcritical bubbles were first discussed in \cite{GleiserKW}. In
\cite{janne} it was shown that phase equilibrium can be reached
provided the transition is weak enough. In these papers, however,
the disappearance of the  subcritical bubbles was  not accounted for.
This happens in two ways: the subcritical bubbles, being unstable
configurations, tend to shrink; the bubbles are also a subject of
constant thermal bombardment so that they may disappear simply because
of thermal noise. The thermalization rate of small-amplitude
configurations near the critical temperature has been estimated in the
EW theory in \cite{per}, where it was found that compared with
typical first order transition times, thermalization is  rather fast.
Kinetics of subcritical bubbles has
 been investigated by Gelmini and
Gleiser \cite{GelminiG}, who found, with a  specific assumption
about the form of the destruction rate due to thermal noise, that
thermal noise becomes subdominant as the Higgs mass is increased. Their
study deals with phase mixing above the critical temperature, and the
possibility that pre-tansitional phenomena might play a role in the
EW phase transition. Phase mixing  has also recently
been simulated numerically by Gleiser \cite{gleiser}
in a 2+1 --dimensional model, with some interesting results. We shall return
to the issue of bubble shrinking by termal noise at the end of the paper.

In the case of a very weak first
order phase transition subcritical bubbles may coalesce
and form large regions of broken phase at and below the critical temperature,
and effectively trigger the phase transition. This happens
if the probability of subcritical
fluctuation is large enough, in which case
 a collection of subcritical bubbles can
form a region of the size of a critical bubble before they have time
to shrink away. This might be called induced nucleation.
Although it is clear  that some dynamics is also involved,
in the present paper we study the formation and growth of subcritical
bubbles in a probabilistic approach. We derive an expression for
the random growth of subcritical regions and compare the rate with
the formation rate of critical bubbles. Interestingly enough, it
turns out that for a large set of the parameter values induced nucleation is
indeed possible,  although for an extremely weak phase transition its
completion is prevented because of the
large size of the critical bubble.

Conventionally two possibilities  for \cb formation in $d + 1$
dimensions is considered. These correspond to the formation rates
\bea {crit}
\Gamma_{d+1}/V &\sim e^{-S_{d+1}[\phi_{cr}]/\hbar },\\
\Gamma_d/V &\sim e^{-\beta S_d[\phi_{cr}]}.
\eea
Eq. (1) describes the situation where the critical bubble is
formed by quantum tunneling, and Eq. (2)
by large over-the-barrier thermal fluctuations.
Note the absence of $\hbar$ in Eq. (2).
$S_d[\phi_{cr}]$ is $d$--dimensional euclidean
action for the critical bubble configuration $\phi_{cr}$.
Such critical bubbles are a result of a single, instantenous fluctuating event
through or over  the potential barrier.
The timescale $\tau_\omega$ of such a \cb
of radius $R_c$ is determined by the thermal mass $m(T)$ and wave number
$k_c \simeq 1/R_c$: $\tau_\omega = 1/\omega = [m(T)^2 + k_c^2]^{-1/2}$.

In a very weakly first order phase transition, however, the two phases
are very much mixed already at the critical temperature, and
thermal processes could create critical bubbles from small, unstable
subcritical bubbles of broken phase. Instead of quantum tunneling or large
fluctuations, the phase transition could be
triggered by small semiclassical thermal fluctuations. This is called
induced nucleation.

Subcritical bubbles will of course disappear very rapidly, but if
their production rate is large enough, the total volume in the broken
phase may actually grow because of the presence of fluctuating
subcritical bubbles.
Our aim is to determine the timescale in which a \cb is thermally
built up from subcritical bubbles by
calculating  the probability for the appearance of a growing
(spherical) subcritical bubble. Following \cite{GelminiG}, we
assume that thermal noise plays a subdominant part, and that
the disappearence of the bubbles can be described in terms
of shrinkage only. As we shall argue, this seems to be a reasonable
approximation in our case.

Let us assume that the \scb is a gaussian, spherically symmetric
configuration with diameter $l$, i.e.
\be {phi}
\phi _l(r) = v(T) e^{- 2 r^2/l^2},
\ee
where subcriticality implies that $l \ll R_c$. At finite temperature we
need the $d$--dimensional action
\be {action}
S_d[\phi_l] =  \int \dx d [\frac 12 (\nabla \phi_l)^2 + V(\phi_l)]
\ee
which essentially determines the formation rate of \scbs per unit volume:
\be {rate}
\Gamma_V(l) \equiv \Gamma[\phi_l]/V \simeq T^{d+1}\left [
{ \beta S_d[\phi_l]\over
2 \pi}\right ]^{d/2}e^{- \beta S_d[\phi_l]}.
\ee

In order to be spesific, we shall deal with a phenomenological potential
for the order parameter $\phi$ suitable for a simple description
of a first order phase transition, given by
\be {pot}
V(\phi) = \frac 12 m(T)^2 \phi^2 - \frac 13 \alpha T \phi^3 + \frac 14 \lambda
\phi^4.
\ee
Most of the analysis takes place at the critical temperature $T_c$ (or
slightly below),  given by the condition
\be {Tmass}
m(T_c)^2 = \frac 29 {\alpha^2 T_c^2\over \lambda}.
\ee
At the critical temperature the non-zero minimum $v(T_c)$ of the
potential reads
\be {vacuum}
v(T_c) = \frac 23 \frac \alpha\lambda T_c
\ee
and the correlation length is
\be {corr}
l_c = {1 \over m(T_c)}.
\ee
We shall let the dimension of the space--time to be a free parameter;
the number of spatial dimensions is denoted  by $d$. Note that
this means that the parameters
$\alpha$ and $\beta$ are, in general, dimensional; only when $d = 3$ they are
dimensionless.

Given the phenomelogical potential \eq{pot}, we can find out the action
for the subcritical configuration \eq{phi} at $T_c$. The result is
\be {scaktio}
\beta S_d = {2\pi \over 9}{\alpha^2 T\over\lambda^2}
\frac d2 [{\sqrt\pi \over 2 m }{l\over l_c}]^{(d - 2)}
\left [1 + {1 - 2 (\frac 23)^{\frac d2} + (\frac 12)^{
\frac d2} \over 2 d}({l\over l_c})^2 \right ].
\ee
The formation rate for subcritical bubbles is then given according
to \eq{rate}.

In the cosmic soup there should exist all sizes  of subcritical bubbles.
We shall, however,
use in our calculations a mean \scb $l = l_c$ with an
associated nuccleation rate $\Gamma_V$. With such an average
subcritical bubble we are able to write down the
probability for a process in which a large number of small
\scbs are created adjacent to each other so that they form a \cb.

Let us first find out how large a part of the space is occupied
by the broken phase
 before the possible formation of critical bubbles, i.e. by subcritical
bubbles
in general. This problem has also been analyzed,
in the case of critical bubbles, in \cite{GuthT} and in \cite{CsernaiK},
however, without the shrinking effects (see also \cite{GelminiG}).
Let the radius of the \scb be $R_{sc}$ and its life--time
\be {nopeus}
t_{sc} =R_{sc}/v~,
\ee
where $v$ is the (average) speed of the shrinking bubble wall.
Let us consider a volume $V$ and, for the moment, discretize the time into
infinitesimal intervalls of lenght $\delta t$. We denote the volume occupied
by the
broken phase at the instance $n \delta t$ ($ n \in \Z_+$) by $V_b(n)$. Between
the times $(n-1) \delta t$  and $n \delta t$ $N_n = \Gamma_V \delta t (V -
V_b(n-1))$ new bubbles of broken phase have been fluctuated,
each having the volume $\pi^{d/2}R_{sc}^d/\Gamma(d/2 + 1)
$. Similarly, at $(n-1)\delta t$ there were
$N_{n-1}$ bubbles but at $t = n\delta t$ their volume has shrunk to
$\pi^{d/2}(R_{sc} - v \delta t)^d/
\Gamma(d/2 + 1)$. This procedure can be continued
until all coexisting bubbles have been counted. Correspondingly,
in the broken phase
$\tilde N = \Gamma_V' \delta t V_b(n-1)$ bubbles of the symmetric phase are
created, and so on. Note that at the critical temperature $\Gamma_V =
 \Gamma_V'$ and $V = V'$. Taking into account both effects, and taking
the limit $\delta t
\goto 0$, we obtain an integral equation for the $V_b$ to $V$ ratio
\be {eqs}
\frac{V_b}{V}\equiv
s(x) = k \integral 0 {{\rm min}\{x,\; 1\} } dy\, [1 - s(x-y)][1-y]^d
       - {k'\over \rho} \integral 0 {{\rm min}\{x,\; \rho \} }
       dy\,  s(x - y)[1 - {y \over \rho}]^d
\ee
with the boundary condition $s(0) = 0$, where $x$ is the scaled time
$x = tv/R_{sc}$, $\rho = (v R'_{sc})/(v' R_{sc})$ and
\be {defk}
k = \Gamma_V {\pi^{d/2} R_{sc}^{d+1}\over \Gamma(d/2 + 1) v}~;~~
k' = \Gamma_V' {\pi^{d/2} (R'_{sc})^{d+1}\over \Gamma(d/2 + 1) v'}~.
\ee
Here $v'$ is the velocity of the unbroken phase bubble wall and $R'_{sc}$
is the radius of the subcritical bubble of symmetric phase in broken phase.
We have not taken into account the thermal noise effect described in
\cite{GelminiG}. It would modify the definitions of
$k$ and $k'$ but here we assume that the resulting changes are small.
 From the integral equation \eq{eqs} the
asymptotic, equilibrium value of $s$ can be easily solved:
\be {asym}
\lim_{x \goto \infty} s(x)\equiv s_{eq} = {k\over d + 1 + k + k'}.
\ee
The equilibrium value at the critical temperature (where $k = k'$) has the
intuitively natural property that $s_{eq} \goto \frac 12$ when $k \goto \infty$
($ \Gamma_V
\goto \infty$). Equation \eq{asym} tells us how large a
part of the spatial volume
is occupied by the broken phase after the system has relaxed
to equilibrium in an
unorganized way so that no bubbles of critical size form. Note also that
when the temperature decreases, $k$ increases and $k'$ decreases so that
$s_{eq}$ increases, too.

The issue at hand is then,
when are the circumstances such that the \scbs can form
a critical bubble? If this happens, the phase transition can be
triggered by subcritical bubbles and
completed by the expansion of the created critical
bubble. Here we have to be slightly under the critical temperature
because no bubble
dynamics is really present exactly at the critical
temperature. We assume that the
\cb is built up layer by layer and consider spherical bubbles only, because
they minimize the number of the \scbs needed for each layer and thus maximize
the probability. Let us say that the radius of a bubble, still subcritical,
which exists at the time $t$, is $R$. In order to that the bubble is really
growing,
in the time $2 R_{sc}/v$ new \scbs have to fill a layer of volume
$V_R = \pi^{d/2} [(R + 2 R_{sc})^d - R^d]/\Gamma(d/2 + 1)$. Because the
process is Poisson distributed, the probability that $N$ bubbles are nucleated
is
\be {pN}
p_N = {1\over N!} \lambda^N e^{-\lambda}
\ee
where
\be {lambda}
\lambda = \Gamma_V V_R {2 R_{sc}\over v} = 2 k [(\frac R{R_{sc}} + 2)^d
- (\frac R{R_{sc}})^d].
\ee
To fill the volume $V_R$ at least $N_R = V_R\Gamma(d/2+1)/(\pi^{d/2}
R_{sc}^d) = (R/{R_{sc}} + 2)^d - (R/{R_{sc}})^d$ bubbles are needed.
The probability that the layer is filled is thus
\be {prob}
p(R) = \sum_{N \geq N_R} p_N.
\ee
With a little algebraic acrobatics the formula \eq{prob} can be cast in the
form
\be {prob2}
p(R) = P(N_R,\, \lambda)
\ee
where
\be {defP}
P(N,\, \lambda) = {1\over \Gamma(N)} \integral 0 \lambda dt\, t^{N-1} e^{-t}
\ee
is a scaled incomplete $\Gamma$--function.

We have derived the probability $p(R)$ that a new layer
of 'elementary' \scbs is formed fast enough
so that the \scb size increases. We have neglected processes where
large \scbs merge. This might be justifiable because a growing
\scb should be a rare event so that the distance between any such
configurations is large. In any case, such processes would only
make the growth rate larger.

When one uses a phenomenological potential of the type
\eq{pot} there arises an extra complication because  in realistic theories
the order parameter (e.g. a Higgs field)
often also has a phase. Then it is not enough that subcritical bubbles
form next to each other but their phases should be correlated, too.
Otherwise there should arise a domain wall between the two bubbles,
and creating such a wall would require extra energy.
In principle, one can easily account also for the phase by introducing
a parameter $\Omega \in [0,\, 1]$ which determines the probability that the
phase  of
the nucleated subcritical bubble is correlated with the phase of the
pre--existing bubble. The probability for formation of $N$ bubbles
is then modified to read
\be {defp'}
p_N' = {1\over N!} (\Omega \lambda)^N e^{-\lambda}
\ee
and thus the probability of layer formation is
\be {probomega}
p(R,\, \Omega) = e^{-(1 - \Omega)\lambda}P(N_R,\, \Omega\lambda).
\ee
The value of $\Omega$ is, however, a dynamical question which we
are unable to address here. When the new \scb
overlaps with the pre--existing bubble, the latter may influence
the phase of the newly forming
bubble and may even force the phases to be correlated. In such a case
$\Omega =1$. Subcritical bubbles, which could be considered as
wave packets of elementary quanta, might also have a velocity relative
to each other. In that case energetics would no longer hinder the removal
of the domain walls, resulting in $\Omega=1$. In what follows, we
shall write down our general expressions for an arbitrary $\Omega$, but
when applying the results, we shall merely assume that $\Omega=1$.

We are now ready to write down the probability that a subcritical bubble
is growing. It is simply the product over all layer probabilities,
\be {totprob}
p(k) = \prod _i p(R_i,\, \Omega) =\exp[{\sum_i \ln P(R_i,\, \Omega)}]
     \equiv \exp[{j({R_c\over R_{sc}},\, \Omega,\, k)}],
\ee
where $R_i = (2 i + 1)R_{sc} $ and $i$ runs from $i = 1$ to
$i = [1/2 + {R_c/2 R_{sc}}] + 1$, i.e. to the value where the bubble
has reached the critical size. For large $R_c/R_{sc}$
the function $j$ can be approximated well by
the integral
\bea {j}
j &\simeq& \frac 12 \integral 0 {R_c} dR\, \ln p(R,\Omega)\\
  &=& \frac 12 \integral 0 {{R_c\over R_{sc}}} du [\ln P(N(u), 2\, k\,
\Omega N(u)) -
2\, k\, (1-\Omega )N(u)],
\eea
where
$N(u) = (u + 2)^d - u^d$. Now $j$ can be computed numerically once the ratio
$R_c/R_{sc}, \ \Omega,\ d$ and $k$ are given.

To proceed further we have to calculate the average formation time $\tau$
and the thickness $\delta$ of a layer. Generally $\delta < 2 R_{sc}$
because the bubble is shrinking simultaneuously as the layer gets filled.
 $N$ bubbles in the
volume $N {\pi^{d/2}} R_{sc}^d/\Gamma(d/2+1)$ are nucleated during the time
$t$
with the probability
\be {pt}
{1\over N!}\left (v k N \frac t{R_{sc}}\right )^N e^{ -v k N \frac t{R_{sc}}}
\ee
and at least $N$ bubbles with probability
\be {totpt}
\sum_{n \geq N} {1\over n!}\left (v k N \frac t{R_{sc}}\right )^n e^{ -v k N
\frac t{R_{sc}}} = P(N, \, v k N \frac t{R_{sc}})~.
\ee
Thus the average filling time of a layer is
\be{layfillt}
\langle t \rangle = \integral {t = 0}{t = \infty} t\, dP(N, \, v k N
\frac t{R_{sc}}) = {R_{sc}\over v k}.
\ee
That is, the average time to form a new layer is ${R_{sc}/(v k)}$. Because
during the same time the bubble shrinks an amount ${R_{sc}/k}$, the bubble
radius after $i^{\rm th}$ layer can be solved recursively. The result is
$R_i = i (2 - {1/ k}) R_{sc}.$ The critical bubble size has been
reached when $i = i_c \equiv (2 - {1/k})^{-1}{R_c/R_{sc}}$ and thus
the formation time of critical bubble is
\be {formt}
\tau = i_c {R_{sc}\over v k} = {R_c/v \over 2 k - 1},
\ee
The average layer thickness is given
by
\be {thickness}
\delta \equiv R_{i + 1} - R_i = (2 - {1\over k})R_{sc}.
\ee

To complete the calculation we have to write down the formula for the
probability $p_c$ that a subcritical
bubble grows up to a critical one. Here we have
to use the layer formation time $\tau$ and thickness $\delta$
as parameters, that is
$\lambda = \Gamma_V\, \langle t \rangle\, {\pi^{d/2}}[(R + \delta)^d -
R^d]/\Gamma(d/2 + 1)$. The calculation results in the expression
\be {formprob}
p_c(k) = e^{j_c},
\ee
where
\be {jc}
j_c = {1\over 2 - {1/ k}}\int_0^{R_c / R_{sc}} du\, [\ln
P(N_c(u), \Omega N_c(u)) - (1 - \Omega )N_c(u)]
\ee
and $N_c(u) = (u + 2 - {1/k})^d - u^d$. Inspection shows that in the case
$\Omega = 1$ the value of $j_c$,
as a function of $k$, converges rapidly to the asymptotic function
$j_c(R_c/R_{sc}) \equiv j_c(R_c/R_{sc}, \Omega = 1, k \rightarrow \infty) $.
A plot of $j_c(R_c/R_{sc})$ is  presented  in Figure 1. For $\Omega = 1$
a good fit for any $k > 1$ is
\be {fit}
j_c({R_c\over R_{sc}}, 1, k) = - 10^{-0.49 + 0.27/k^{1.15}}\, \left (
{R_c\over R_{sc}}\right )^{1.01 + 0.008/(k - 0.2)^{1.20}}.
\ee
For decreasing $\Omega\ \ j_c$ increases, however, rapidly even for as small
ratios as ${R_c/ R_{sc} = 10}$. In Figure 2 we
demonstrate this behaviour by presenting $j_c$ as a function
of $\Omega$ for $k \rightarrow \infty$ and $R_c/R_{sc} = 10$.

We may now apply the rate \eq{formrate} to cosmology.
Because \eq{formrate} states that
$\Gamma_{V, f} \propto e^{- (\beta S_d[\phi_l] - j_c)}$,
we have to compare the combination $\beta S_d[\phi_l] - j_c \equiv S^{eff}$
with the Hubble rate. An induced critical bubble is obtained when
\be{comp}
S^{eff} \approx \ln \left ({M_{Pl}\over T}\right )^4~.
\ee

The probability $p_c(R)$ is, in general, very small.
The number of \scbs during the
phase transition, i.e. the number of possible seeds, can,
however, be very large.  In this case the \cb
formation rate is given by
\be {formrate}
\Gamma_{V,\, f} = p_c(k)\, \Gamma_V.
\ee
where $\Gamma_V$ can be computed by using \eq{scaktio}.
To get some idea of the magnitudes of the  parameters needed for
 induced nucleation, we compute the rate \eq{formrate}
in the thin wall approximation using the potential \eq{pot} and assume that
$v$ is of the order of unity.

First we need the size of the critical bubble.
Assuming that there is only little supercooling, the bounce action
can be written as \cite{janne}
\be{bounce}
S/T=\frac\alpha{\lambda^{3/2}}{2^{9/2}\pi\over 3^5}{\bar{\lambda}^{3/2}\over
(\bar{\lambda}-1)^2}\simeq 150~,
\ee
where
\be{barlam}
\bar{\lambda}\simeq 1-0.0442{\alpha^{1/2}\over \lambda^{3/2}}\equiv
1-\delta~.
\ee
Small supercooling, or $1-\bar\lambda\ll 1$ thus implies that
$\alpha\ll 500\lambda^3$.
Solving for $\bar\lambda$ yields the transition temperature  $T_f$.
It then follows that at the transition temperature
\be{massa}
m^2(T_f)={2\alpha^2\over 9\lambda}(1-\delta)T_f^2~.
\ee
Expanding the potential at the broken minimum $\phi=v(T)=
\alpha T(1+\sqrt{1-8\bar\lambda/9})/2\lambda$ we find
\be{epsiloni}
-\epsilon\equiv V(v,T_f)=\frac 16 m^2(T_f)v^2-\frac {1}{12}\lambda v^4
=-0.00218 \left({\alpha\over\lambda}\right)^{9/2}+{\cal O}(\delta^2)~.
\ee
The  height of the barrier is situated at $\phi_{\rm max}\simeq
v/2$ with $V(\phi_{\rm max},T_c)\equiv V_{\rm max}
=\alpha^4T_c^4/(144\lambda^3)$. As $T_c\simeq T_f$ we may conclude
that the thin wall approximation is valid if $-\epsilon/V_{\rm max}
=0.314\alpha^{1/2}/\lambda^{3/2}\ll 1$, or $\alpha\ll 10\lambda^3$,
which is in accordance with our assumption of small supercooling.

To get the size of the critical bubble we still need the surface
tension. One easily finds
\be{surface}
\sigma=\int_0^\infty d\phi\sqrt{2V(T_c)}={2\sqrt{2}\alpha^3
\over 91\lambda^{5/2}}T_c^3~.
\ee
Thus we finally obtain
\be{ratio}
{R_c\over R_{sc}}={2\sigma\over\epsilon R_{sc}}
\simeq 26.9{\lambda^{3/2}\over\alpha^{1/2}}~.
\ee
Here $R_{sc}\simeq 2/m(T_f)$.
Since $\alpha\ll 10\lambda^3$ we see that ${R_c/ R_{sc}}\gg 1$ as it should.

Combining \eq{scaktio}, \eq{fit}, \eq{comp} and \eq{ratio} at the critical
temperature we obtain a condition for inducing a critical bubble:
\be{res}
2.06 {\alpha\over \lambda^{3/2}} + 9.10
\left ( {\lambda^3\over \alpha}\right )^{0.506} \lsim 150.
\ee
This implies $\alpha \gsim 4.0\times 10^{-3}\lambda^3$, that is,
perhaps surprisingly,
induced nucleation does not work for a too weak transition.
The result can be
interpreted so that in the range $1\lsim \lambda^3/\alpha \lsim 250$ where the
phase transition is very weak, it is triggered by subcritical bubble formation
immediately below the critical temperature.
If the phase transition is weaker, the critical bubble radius becomes
too large to be induced by subcritical bubble growth.
On the other hand, whenever
$\lambda^3/\alpha \lsim 1$ the phase transition is not weak any longer:
the radius of a critical bubble decreases and bubble dynamics becomes
inportant so that our calculation loses its validity. The correlation
problem connected to the parameter $\Omega$ also  becomes
 more important when the
``seed bubble'' is of same size as the newly produced ones.

To appreciate the weakness of the transition required of induced nucleation,
we may consider the 2-loop result for the electroweak effective
potential calculated in \cite{studies2}. Their result
for $M_H=87$ GeV is well fitted
by the potential of the type \eq{pot} with $\alpha=0.048$ and $\lambda
=0.061$. Thus in this case $\alpha\simeq 210\lambda^3$
is well outside of the induced nucleation range. As for larger Higgs masses
$\alpha$ is likely to be smaller, it is conceivable that induced nucleation
could play a role in EW phase transition. We should however emphasize that
the result \eq{res}, and the subsequent estimates, should be considered
as the most optimistic cases. Dynamics that correlates phases, and other
issues related to the velocity of the subcritical bubbles or the
surface tension between the bubbles, are likely to make induced nucleation
more difficult, not facilitate it.

Regarding thermal noise, we have compared the thermalization rate of small
amplitude configurations in the EW theory \cite{per}, given by $\Gamma \simeq
10^{-2} T$, with the formation time of an induced critical bubble,
\eq{formt}. We find that in the region where the thin wall approximation
is valid, thermalization does not occur. This seems to indicate that thermal
noise does not play an important role in induced nucleation.

It would be of interest to  compare the present approach with the
results of Gleiser \cite{gleiser}, although it
is not obvious to us how to do it. The problem appears to be a technical
one only, though.  Gleiser also employed a potential of the type
\eq{pot} (in $d=2$) but simulated the heat bath by white noise which
introduced an additional parameter to the problem. It
is not clear how the parameters in his simulation are related
to the parameters (including $T_c$) here.

\newpage

\newpage
%
\noindent {\Large{\bf Figure captions}}
\vskip .5truecm
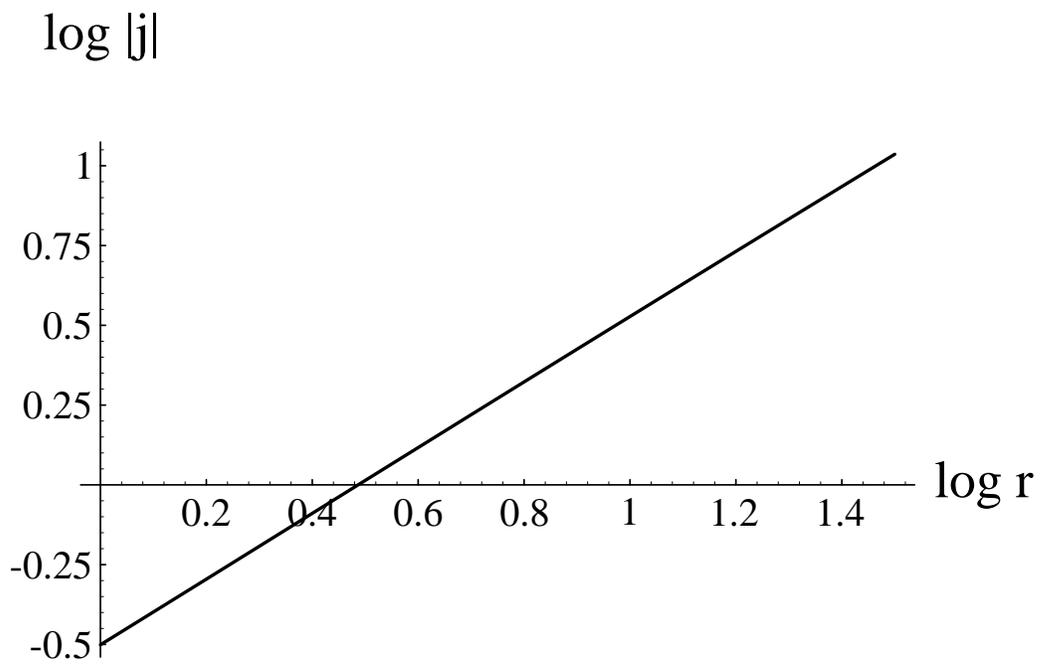
\noindent Figure 1. The probability function $j(r,
\Omega = 1)$ defined in \eq{j}
in the asymptotic limit $ k \rightarrow \infty$
as a function of the ratio of critical and subcritical bubble radii
$r = R_c/R_{sc}$.
\vskip .5truecm
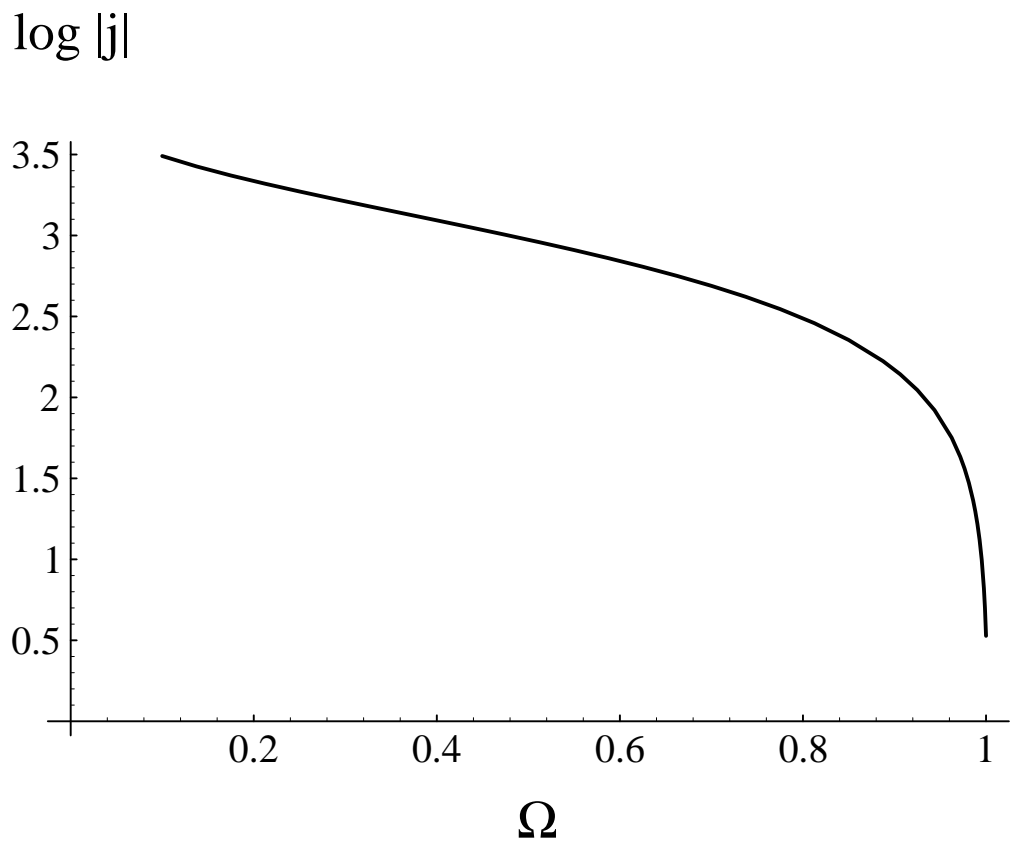
\noindent Figure 2. The function $j(r,
\Omega)$ in the asymptotic limit $ k \rightarrow \infty$
as a function of $\Omega$ for the fixed ratio $r = R_c/R_{sc} = 10$.

\begin{thebibliography}{X}
%
\bibitem{bg} For a review, see A.G.\ Cohen, D.B.\ Kaplan and A.E.\ Nelson,
Ann.\ Rev.\ Nucl.\ Part.\ Phys.\ {\bf 43} (1993) 27.
\bibitem{studies1} K. Kajantie, K. Rummukainen and M. Shaposnikov,
Nucl.~Phys.~{\bf 407} (1993) 356;
K. Farakos et al., Preprint CERN-TH.7244/94.
\bibitem{studies2} W.\ Buchm\" uller et al.,
 Preprint DESY-93-021; Z.\ Fodor et
al., Preprint DESY-94-088.
\bibitem{FodorH} Z.\ Fodor, A.\ Hebeker, Preprint DESY-94-025.
\bibitem{GleiserKW} M.\ Gleiser, E.\ Kolb and R.\ Watkins, Nucl.\ Phys. {\bf
B364} (1991) 411.
\bibitem{janne} K.\ Enqvist et. al., Phys. Rev. {\bf D45} (1992) 3415.
\bibitem{per} P.~Elmfors, K.~Enqvist and I.~Vilja, Nucl. Phys.
{\bf 412} (1994) 459.
\bibitem{GelminiG} G.\ Gelmini and M.\ Gleiser, Nucl.\ Phys. {\bf B419}
(1994) 129.
\bibitem{gleiser} M.\ Gleiser, Preprint DART-HEP-94/01.
\bibitem{GuthT} A.H.\ Guth and S.-N.\ Tye, Phys.\ Rev.\ Lett.\ {\bf 44} (1980)
631, 963 (erratum). See also A.H.\ Guth and E.J.\ Weinberg, Phys.\ Rev.\
{\bf D23} (1981) 876.
\bibitem{CsernaiK} L.P.\ Csernai and J.I.\ Kapusta, Phys.\ Rev.\ Lett.\ {\bf
69}
(1992) 737.
\end{thebibliography}
\end{document}